\pdfoutput=1

\documentclass[11pt]{article}

\newcommand{\minisection}[1]{
    \vspace{1pt}\noindent\textbf{#1}
}

\usepackage[final]{acl}
\usepackage{booktabs}
\usepackage{times}
\usepackage{latexsym}
\usepackage{listings}
\usepackage{xcolor}
\usepackage{enumitem}
\usepackage{amsmath}

\usepackage[T1]{fontenc}

\usepackage[utf8]{inputenc}

\usepackage{microtype}

\usepackage{inconsolata}

\usepackage{graphicx}
\usepackage{subcaption}

\title{REALM: A Dataset of Real-World LLM Use Cases}

\author{
  Jingwen Cheng\textsuperscript{1}, 
  Kshitish Ghate\textsuperscript{1}, 
  Wenyue Hua\textsuperscript{2}, 
  William Yang Wang\textsuperscript{2}, 
  Hong Shen\textsuperscript{1}, 
  Fei Fang\textsuperscript{1} \\
  \textsuperscript{1}Carnegie Mellon University, Pittsburgh, USA \\
  \textsuperscript{2}University of California, Santa Barbara, USA \\
  \texttt{\{chengjw21\}@gmail.com} \\
  \texttt{\{kghate, hongs, feif\}@andrew.cmu.edu} \\
  \texttt{\{wenyuehua, william\}@cs.ucsb.edu}
}

\begin{document}
\maketitle
\begin{abstract}
Large Language Models (LLMs), such as the GPT series, have driven significant industrial applications, leading to economic and societal transformations. However, a comprehensive understanding of their real-world applications remains limited. To address this, we introduce \textbf{\textsc{REALM}}, a dataset of over 94,000 LLM use cases collected from Reddit and news articles. \textbf{REALM} captures two key dimensions: the diverse applications of LLMs and the demographics of their users. It categorizes LLM applications and explores how users’ occupations relate to the types of applications they use. By integrating real-world data, \textbf{\textsc{REALM}} offers insights into LLM adoption across different domains, providing a foundation for future research on their evolving societal roles. An interactive dashboard (\url{https://realm-e7682.web.app/}) is provided for easy exploration of the dataset.
\end{abstract}

\section{Introduction}
Large Language Models (LLMs) have revolutionized AI with advanced natural language understanding and reasoning capabilities \cite{guo2025deepseek, jaech2024openai, team2024gemini}. These advancements have spurred new applications and transformed workflows across various industries. As LLM capabilities expand, researchers in fields like social science, economics, and human-computer interaction (HCI) are examining their social implications, anticipating both benefits and disruptions, and addressing risks tied to their rapid adoption.

Currently, there remains limited understanding of how LLMs are deployed and utilized in real-world scenarios despite widespread adoption.
Existing studies generally fall into two primary categories. The first uses occupation-activity typologies to measure LLM’s influence on various tasks, aiming to assess task-exposure rates and impact on occupations \citep{Abril07, Felten2018, Brynjolfsson2018, Webb2020}. However, these studies often lack empirical grounding, weakening their validity. The second category analyzes large datasets from social media and other public sources to evaluate sentiment and trends around LLM applications \citep{Leiter2023, Miyazaki2024, Koonchanok2024, Naing2024}, but these studies are hindered by a lack of comprehensive taxonomies and rely on simplistic methods, such as keyword filtering, leading to low precision and recall. While some industry research (e.g., from LinkedIn, Anthropic, Goldman Sachs) on LLM impact is available, it is often not publicly accessible or prohibitively costly, limiting its use in academic research.

To conduct an empirical study with a comprehensive taxonomy, we present a new dataset \textbf{Re}al-world \textbf{A}pplications of Large \textbf{L}anguage \textbf{M}odels (\textbf{\textsc{REALM}}), documenting real-world use cases of LLMs. \textbf{\textsc{REALM}} comprises over 94,000 instances, including 15,000 sourced from Reddit discussions and 79,000 from news articles, spanning from June 2020 (the release date of GPT-3) to December 2024. Our analysis focuses on two primary dimensions: the objectives of LLM use case based on a comprehensive taxonomy \cite{Theofanos2024AITaxonomy}, and the occupations of current or potential end-users, categorized using the O*NET database \cite{handel2016net}. Our contributions are threefold: (1) we employ a systematic and reliable taxonomy to link use cases to occupations (2) we provide initial statistical analyses derived from the dataset, demonstrating its potential for analyzing trends, informing policies, and supporting cross-domain research (3) we provide accessible via an interactive dashboard and an API for structured data processing. 

\section{Related Work}
Studying the impact of LLM-based applications has emerged as a prominent research topic, including both structured analyses based on existing occupation-activity taxonomies and empirical investigations utilizing social media big data. 

Some research examines the societal and economic impacts of LLMs through structured taxonomies, employing either GPT-based or human annotations to assess potential LLM usage. For example, \citet{Abril07} estimates that 19\% of U.S. workers could have 50\% or more of their tasks exposed to LLMs using a task-exposure rubric derived from the O*NET database. This study builds on a broader body of work that quantifies AI exposure rates \cite{Felten2018, Webb2020, Brynjolfsson2018}. However, these analyses often rely on predefined rubrics and expert opinions for annotation, which can introduce biases and fail to reflect real-world dynamics.

In contrast to structured taxonomy-based approaches, empirical research leverages social media data to explore real-world applications of LLMs \cite{Naing2024}. For instance, \citet{Leiter2023} analyzes sentiment trends and employs topic modeling to investigate the impacts of LLMs across domains such as education and technology, but utilizing a highly simplified taxonomy. \citet{Miyazaki2024} and \citet{Koonchanok2024} incorporate occupation-level analyses based on O*NET and Indeed to examine how different professions perceive and interact with LLMs. \citet{Miyazaki2024} further analyzes user-provided prompts and applies topic modeling to understand user interactions with LLMs. However, these studies often depend on keyword filtering or hashtag-based data collection methods, which are imprecise and fail to provide a comprehensive taxonomy due to the necessity of omitting many keywords that may have multiple meanings. Beyond social media, other research has examined LLM adoption through alternative data sources with varying focuses, including scientific literature \cite{Gao2024}, large-scale user interaction logs with LLMs \cite{Zheng2023, zhao2024wildchat}, and proprietary usage data from commercial LLM deployments \cite{handa2025economic}. The latter offers a comprehensive analysis of millions of conversations with Claude, revealing correlations with user occupations, salaries, and job zones. However, their data is not publicly available, and the results are limited to a relatively short time frame.


Building upon the aforementioned works, \textbf{\textsc{REALM}} introduce a taxonomy-driven approach coupled with a semantically rich filtering process. We systematically link LLM applications to specific occupations and use cases using both social media and news platforms with high-performing data collection and cleaning pipeline. Unlike prior studies based on proprietary or short-term data, REALM continuously aggregates public content across diverse domains over an extended time span, enabling a precise and nuanced understanding of how different professional sectors are integrating LLMs into their workflows, offering a robust foundation for future research.

\section{Dataset Construction}
\begin{figure}[t]
  \centering
  \includegraphics[width=\columnwidth]{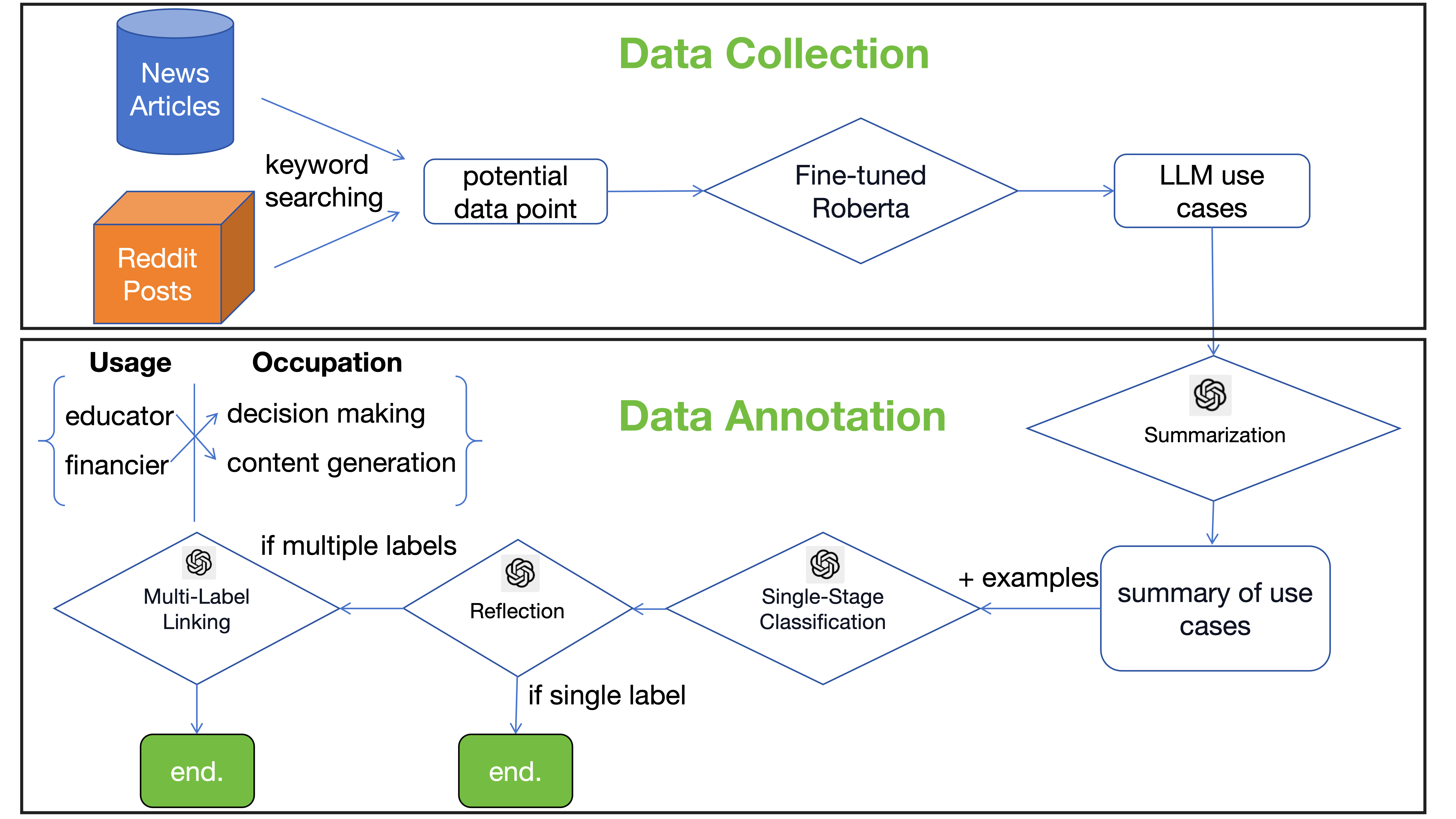}
  \vspace{-10pt}
  \caption{REALM Dataset Construction Pipeline}
  \vspace{-10pt}
  \label{fig:pipeline}
\end{figure}

\vspace{-5pt}
The construction of the \textbf{REALM} follows a structured pipeline for collecting and processing high-quality data, as illustrated in Figure \ref{fig:pipeline}. A dedicated dashboard \textbf{\url{https://realm-e7682.web.app/}} is also built to facilitate data exploration and visualization. 
Construction details are presented below.

\begin{table*}[t]
  \centering
  \small
  \resizebox{0.95\textwidth}{!}{ 
  \begin{tabular}{ccccc|cccc}
    \toprule
    \multicolumn{4}{c}{\textbf{LLM Use Categorization}} & & \multicolumn{4}{c}{\textbf{Occupation Categorization}} \\
    \cmidrule{1-4} \cmidrule{6-9}
    Class & Precision & Recall & F1-Score & & Class & Precision & Recall & F1-Score \\
    \midrule
    Content Creation & 0.75 & 0.83 & 0.79 & & Management & 0.87 & 0.62 & 0.72 \\
    Content Summarization & 0.70 & 0.88 & 0.78 & & Business and Financial Operations (Bus \& Finance) & 0.82 & 0.70 & 0.76 \\
    Decision Making & 0.71 & 0.74 & 0.72 & & Computer and Mathematical (Comp \& Math) & 0.72 & 0.86 & 0.78 \\
    Detection & 0.68 & 0.73 & 0.70 & & Architecture and Engineering (Arch \& Eng) & 0.80 & 0.75 & 0.77 \\
    Digital Assistants & 0.69 & 0.79 & 0.74 & & Life, Physical, and Social Science (Life \& Science) & 0.73 & 0.81 & 0.77 \\
    Discovery & 0.74 & 0.78 & 0.76 & & Community and Social Service (Comm \& Social) & 0.69 & 0.82 & 0.75 \\
    Image Analysis & 0.80 & 0.85 & 0.82 & & Legal & 0.78 & 0.85 & 0.81 \\
    Information Retrieval & 0.79 & 0.78 & 0.78 & & Educational Instruction and Library (Edu \& Library) & 0.80 & 0.79 & 0.79 \\
    Personalization & 0.70 & 0.73 & 0.71 & & Arts, Design, Entertainment, Sports, and Media (Art \& Media) & 0.81 & 0.77 & 0.79 \\
    Prediction & 0.78 & 0.83 & 0.80 & & Healthcare Practitioners and Support (Health \& Care) & 0.70 & 0.79 & 0.74 \\
    Process Automation & 0.68 & 0.89 & 0.77 & & Sales and Related (Sales) & 0.77 & 0.75 & 0.76 \\
    Recommendation & 0.72 & 0.74 & 0.73 & & Office and Administrative Support (Office \& Admin) & 0.76 & 0.82 & 0.74 \\
    Robotic Automation & 0.79 & 0.83 & 0.81 & & Production & 0.73 & 0.81 & 0.77 \\
    Vehicular Automation & 0.82 & 0.86 & 0.84 & & Others & 0.64 & 0.80 & 0.71 \\
    \bottomrule
  \end{tabular}
  }
  \vspace{-5pt}
  \caption{Performances of our pipeline for LLM Use categorization and Occupation categorization tasks separately.}
  \vspace{-10pt}
  \label{tab:commands}
\end{table*}

\paragraph{Data Collection} 
\label{sec: data_collection}
Data in \textbf{\textsc{REALM}} is sourced from Reddit (via Academic Torrents\footnote{Academic Torrents is a distributed system for sharing academic datasets and resources. More information is available at \url{https://academictorrents.com}}) and news articles (retrieved using NewsAPI\footnote{\url{https://newsapi.org/}}) spanning from June 2020 (when GPT-3 was released) to December 2024. This process narrows down billions of datapoints to a curated dataset of 94,000 entries, containing 15,000 Reddit posts and 79,000 news articles. The real-time, user-generated content from Reddit provides diverse perspectives, while NewsAPI offers professionally curated articles, ensuring diversity.

The data collection process begins with keyword-based extraction to isolate LLM-related content, using a comprehensive list of model names obtained from GitHub\footnote{\url{https://github.com/Hannibal046/Awesome-LLM}}. Then we perform essential data cleaning to ensure the quality of the collected content. This includes removing duplicates, unusable tweets, stop words, and non-textual elements, ensuring the data is well-prepared for further processing. We employ a fine-tuned RoBERTa model to filter irrevelant data, ensuring relevance by identifying descriptions of LLM use cases or potential applications. The details of the RoBERTa fine-tuning process are provided in Appendix~\ref{appendix:roberta-finetuning}. This filtering process prioritizes recall over precision to include all pertinent data, achieving high recall scores of 0.94 for news articles and 0.95 for Reddit posts after human validation. Following the initial filtering, a four-stage annotation pipeline, detailed in Section \ref{sec:pipeline}, is applied to further refine the dataset by enhancing precision.

\paragraph{Data Annotation}
\label{sec:pipeline}
The annotation for \textbf{REALM} involves labeling LLM use case types and occupations. We use (1) LLM Use taxonomy\citep{Theofanos2024AITaxonomy} and (2) the Occupation taxonomy. Further details on the taxonomies are provided in Appendix \ref{sec:taxonomy selection}. To simplify analysis, 9 occupation groups with less than 1\% frequency are merged into ``Others'', and healthcare-related roles are consolidated into ``Healthcare'' \cite{Koonchanok2024} to obtain a total of 14 occupations. We utilize a four-module annotation pipeline using GPT-4o-mini to streamline the process, consisting of Summarization, Classification, Reflection and Multi-Label Linking module.
Further information about these modules can be found in Appendix \ref{sec:4-modules}. Table \ref{tab:commands} presents the precision, recall, and F1 score of the four-stage pipeline on a validation dataset of 1000 data points.

\paragraph{Performance Validation}
To evaluate the performance of the whole pipeline, two expert annotators labeled a total of 3,000 data points.\ref{tab:commands} 
See Appendix \ref{sec:annotation details} for more information.

\begin{figure*}[!t]
  \centering
  \begin{subfigure}[b]{0.469\textwidth} 
    \includegraphics[width=\textwidth]{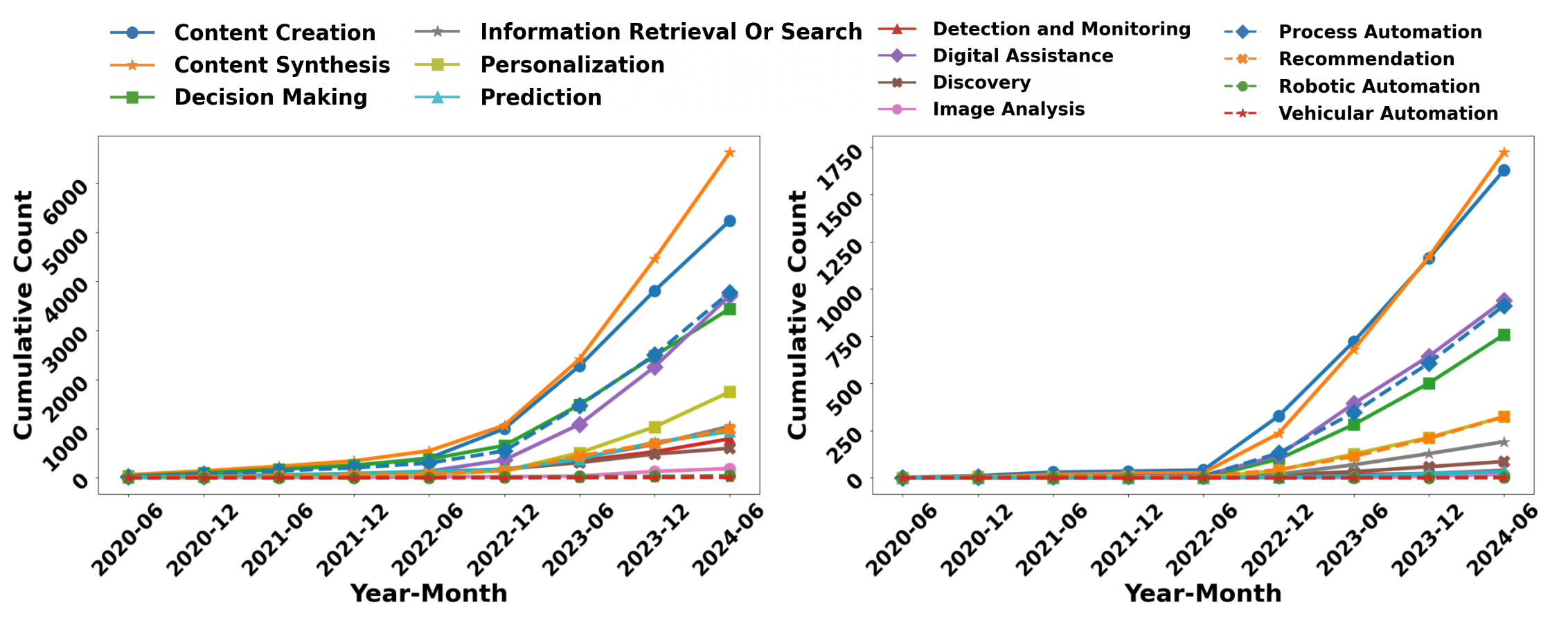}  
    \captionsetup{font=scriptsize}
    \caption{News Article(left) and Reddit Posts(right) Count by LLM Use}
    \label{fig:image1}
  \end{subfigure}%
  \begin{subfigure}[b]{0.494\textwidth}
    \includegraphics[width=\textwidth]{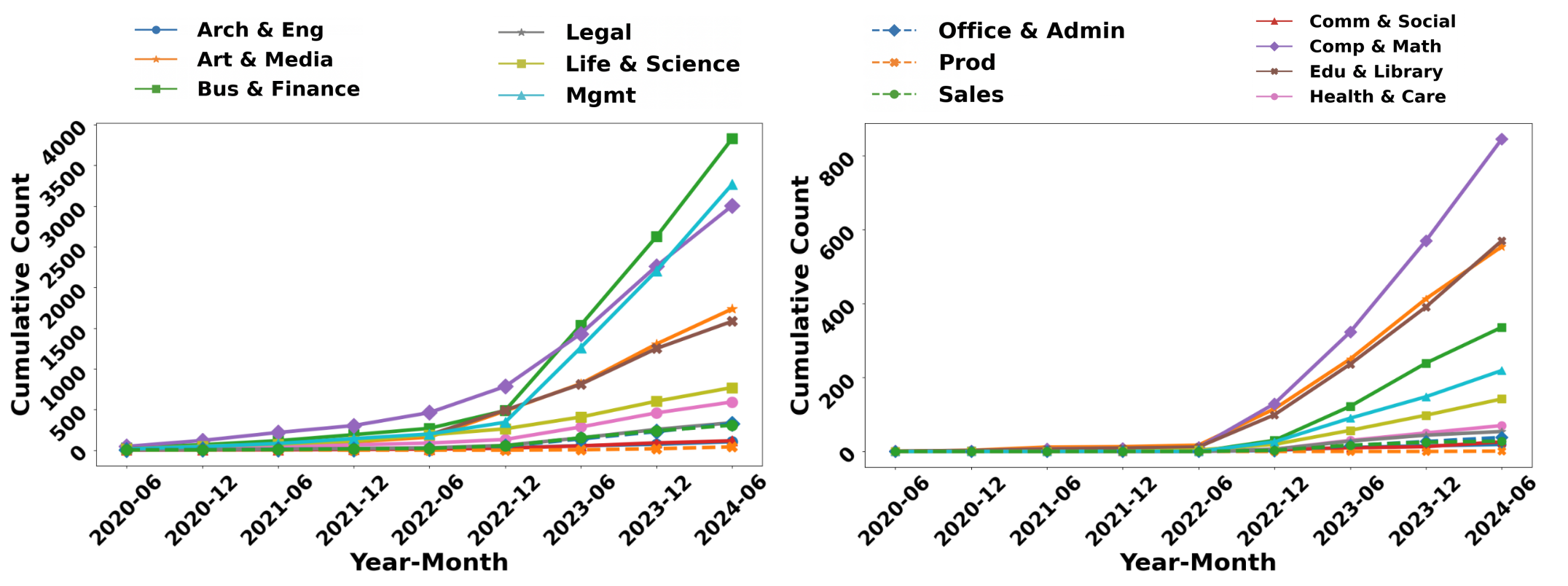} 
    \captionsetup{font=scriptsize}
    \caption{News Article(left) and Reddit Posts(right) Count by Occupation}
    \label{fig:image2}
  \end{subfigure}
  
  \captionsetup{font=footnotesize}  
  \begin{subfigure}[b]{0.4\textwidth}  
    \hspace{-30pt}
    \includegraphics[width=\textwidth]{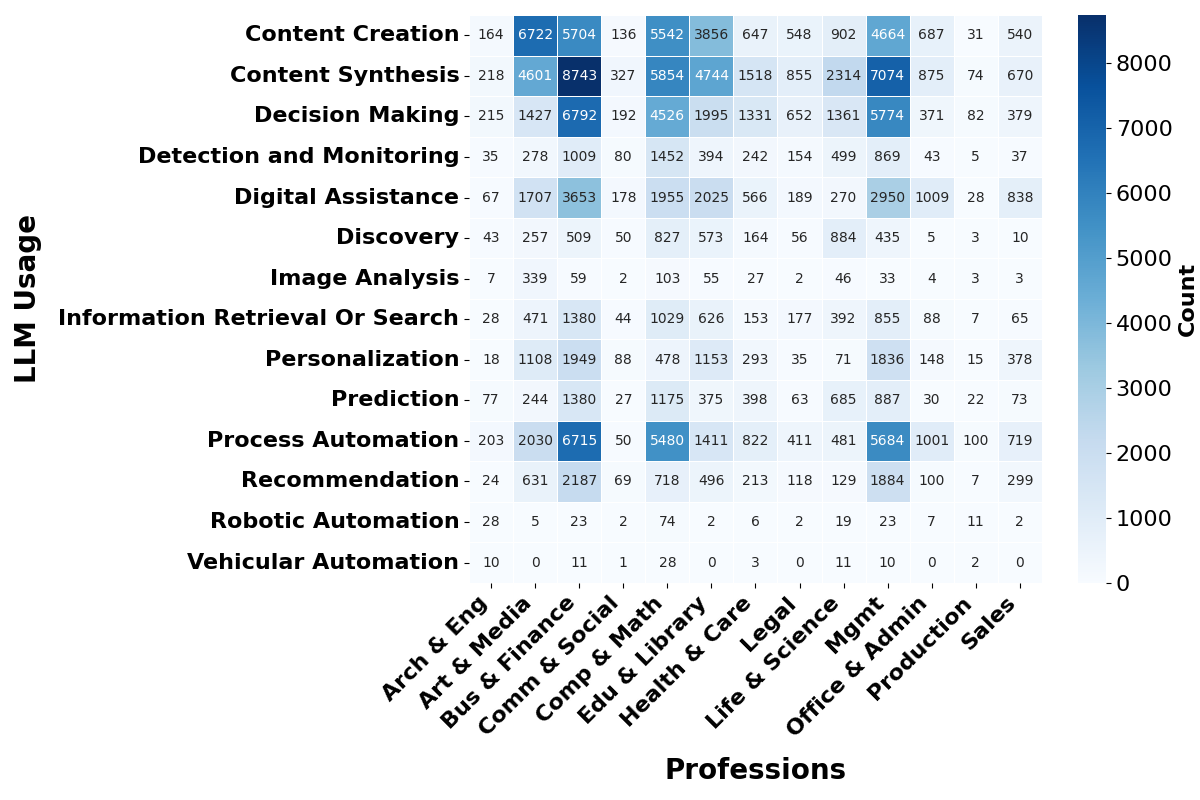}  
    \caption{LLM Use for Different Professions: News Articles}
    \label{fig:image3}
  \end{subfigure}%
  \begin{subfigure}[b]{0.4\textwidth}  
    \hspace{15pt}
    \includegraphics[width=\textwidth]{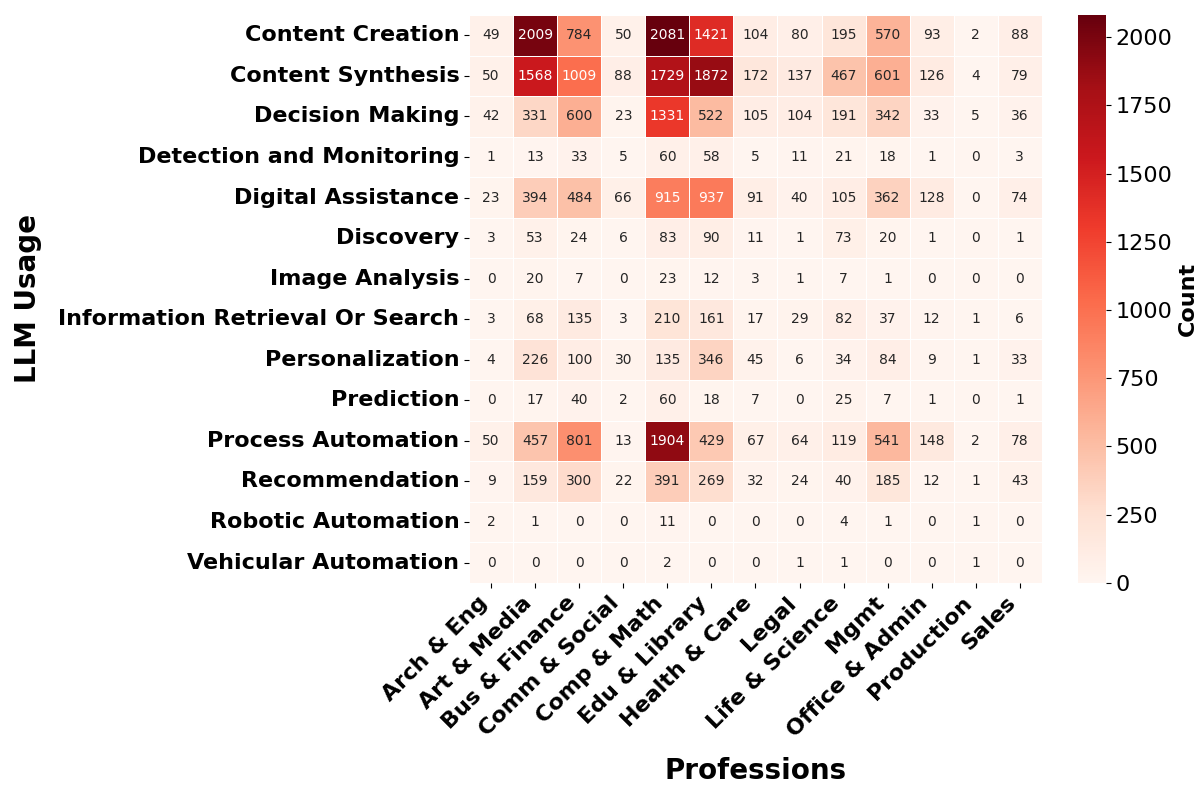}  
    \caption{LLM Use for Different Professions: Reddit Posts}
    \label{fig:image4}
  \end{subfigure}
  \caption{Visualization of LLM Mentions Across Usage and end-user Professions}
  \label{fig:sidebysidefour}
\end{figure*}

\vspace{-5pt}
\section{Analysis of REALM Dataset}
\vspace{-5pt}

In this section, we present preliminary analysis results derived from the \textbf{\textsc{REALM}} dataset. As depicted in Figures \ref{fig:image1} and \ref{fig:image2}, the adoption and application of LLMs experienced a significant surge from between 2022 and 2023, coinciding with the release of GPT-3.5 in November 2022.

Figure \ref{fig:image1} illustrates the number of LLM use cases categorized by application type within News and Reddit sources, respectively. These figures reveal that Content Creation and Content Synthesis are the most prevalent LLM use cases, including writing assistance, summarization, and code generation. Additionally, applications related to Decision-making, Digital Assistance, and Process Automation have experienced substantial growth.

The Decision-making category includes activities such as recommending business strategies, optimizing marketing campaigns, and suggesting financial investment strategies. Process Automation involves streamlining routine and repetitive tasks. The increasing interest in these use cases indicates that users are perceiving LLMs as intelligent and reliable collaborators, thereby building trust in their capabilities to help daily work. This trend signifies that LLMs are evolving beyond their traditional role as generative AI, actively supporting and enhancing industrial workflows.
However, certain areas such as Robotics Automation, Recommendation Systems, and Vehicular Automation have not yet extensively integrated LLM technologies. 

Figure \ref{fig:image2} illustrates the utilization of LLMs across various occupations. In News sources, discussion were initially dominated by professionals in Computer Science and Mathematics. However, over time, people in Business and Management have assumed a more prominent role in these discussions. In contrast, as Reddit has a more technically inclined user base, its discussion is consistently dominated by individuals from computer-related fields. But there has been a growing interest in LLMs among professionals in Art \& Media, Product Development and Education.

Figures \ref{fig:image3} and \ref{fig:image4} demonstrate varying levels of applicability and adoption of LLM technologies across different professional domains: individuals in Art, Business, Computer Science, and Management are the most engaged with LLMs across both News and Reddit sources. Their primary activities include content creation, content synthesis, decision-making, digital assistance, and process automation. Additionally, professionals in Computer Science, Mathematics, Physical Sciences, and Social Sciences are notably involved in discovery-oriented tasks, highlighting the potential of LLMs to advance scientific research. In contrast, occupations in the legal and sales sectors, as well as roles centered on manual or operational tasks (food preparation, transportation, and farming) exhibit limited engagement with LLMs. 

\paragraph{Debiasing}\label{sec:debiasing}
To address potential biases arising from the uneven distribution of occupational groups, we conducted a debiasing analysis to normalize raw LLM use case statistics. Certain fields, such as Arts or Business, may dominate the dataset due to higher overall content volume, artificially inflating their apparent LLM adoption.

We addressed this by applying a normalization procedure. We randomly sampled 1,000 data points from both the news and Reddit corpora and manually annotated each item with its associated occupational group. Based on these annotations, we estimated the discussion distribution across occupations: for each group $i$, we computed its discussion proportion $D_i$, defined as the fraction of sampled items belonging to group $i$. This captures how prevalent each group is in the overall corpus, independent of LLM-specific content.

Next, we normalized the number of LLM use cases associated with each occupational group by dividing by $D_i$, resulting in what we refer to as the \textit{LLM exposure rate}:
\[
\text{ExposureRate}_i = \frac{\text{NumOfUseCases}_i}{D_i}
\]
This metric reflects the relative concentration of LLM-related content within each group, adjusted for how frequently the group appears in the dataset overall. By using this normalization, we aim to distinguish true differences in LLM adoption from artifacts of dataset imbalance. 

\begin{figure}[t]
  \centering

  \begin{subfigure}{\linewidth}
    \centering
    \includegraphics[width=1.0\linewidth]{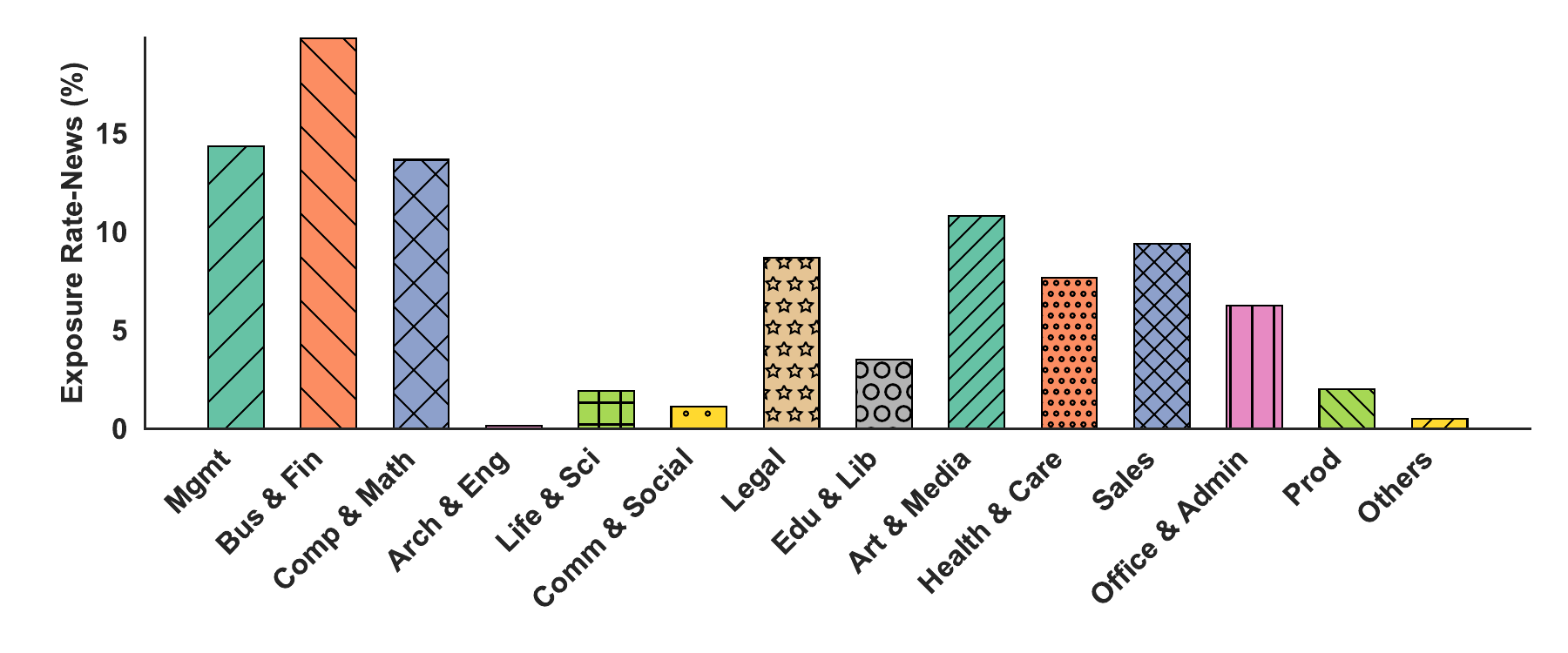}
    \caption{Exposure Rate of News Data.}
    \label{fig:news_exposure}
  \end{subfigure}

  \vspace{0.8em}

  \begin{subfigure}{\linewidth}
    \centering
    \includegraphics[width=1.0\linewidth]{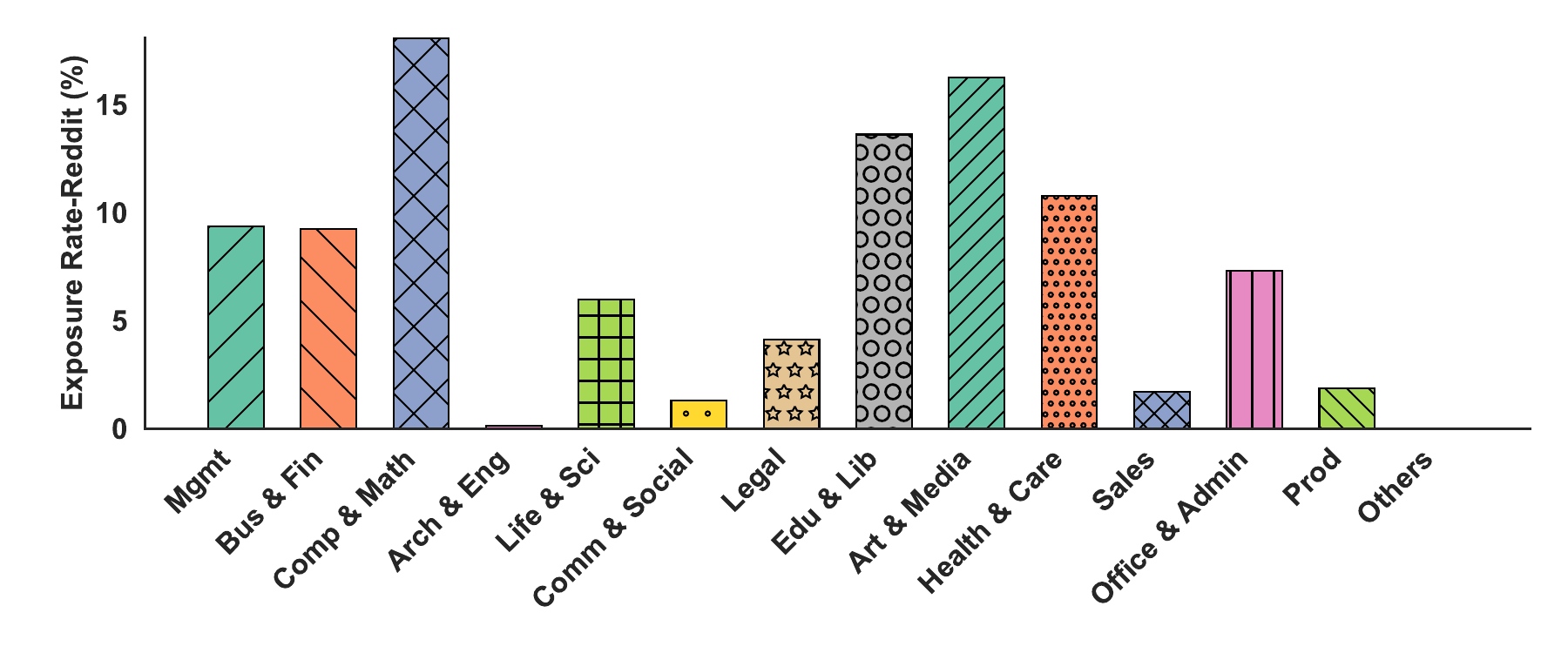}
    \caption{Exposure Rate of Reddit Data.}
    \label{fig:reddit_exposure}
  \end{subfigure}

  \vspace{0.8em}

  \begin{subfigure}{\linewidth}
    \centering
    \includegraphics[width=1.0\linewidth]{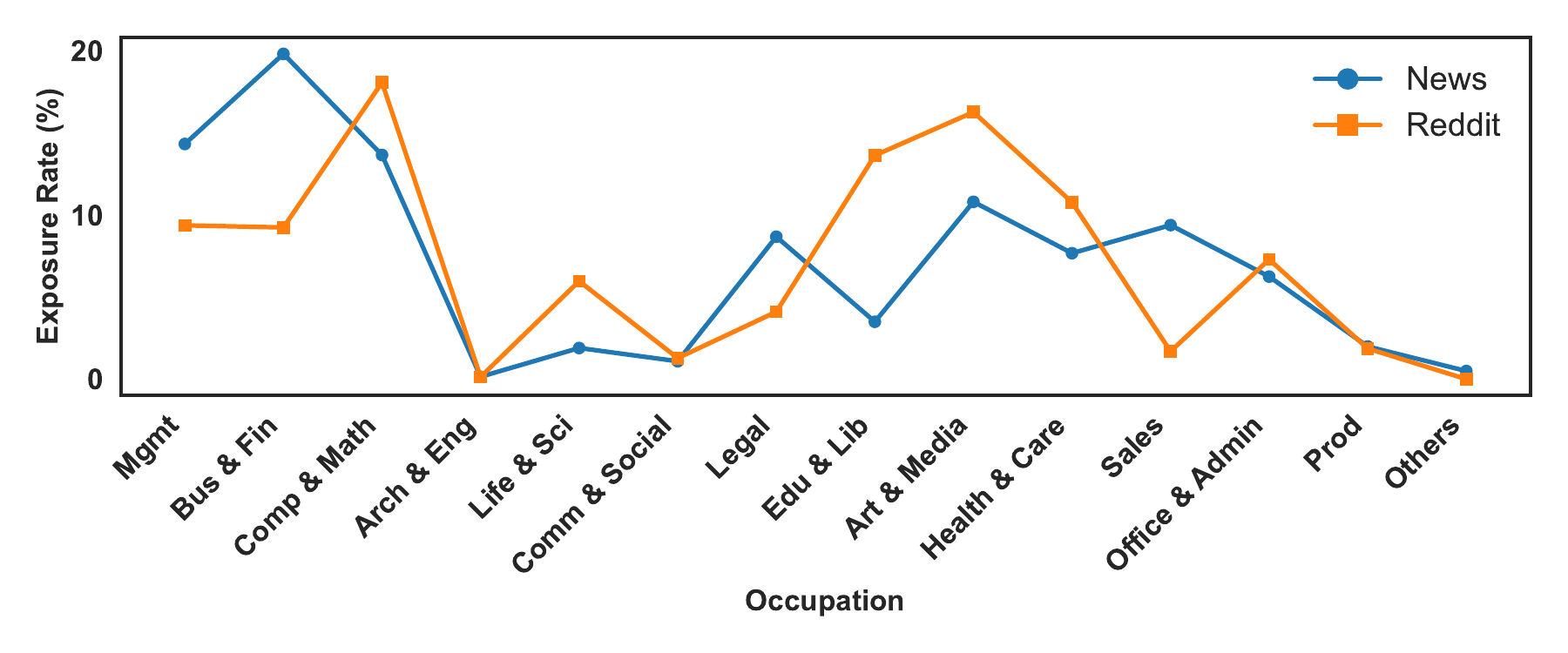}
    \caption{Comparison of Exposure Rate between News and Reddit Data.}
    \label{fig:exposure_comparison}
  \end{subfigure}

  \caption{Visualization of Exposure Rate after debiasing.}
  \label{fig:llm_combined}
\end{figure}

After debiasing (Figure~\ref{fig:llm_combined}), the exposure rates of previously dominant fields such as \textit{Business}, \textit{Management}, and \textit{Computer Science} decrease markedly relative to their raw use case counts (Figure~\ref{fig:reddit_exposure}). This indicates that their initial prominence was primarily driven by the sheer volume of content associated with these occupations, rather than particularly high LLM adoption. In contrast, domains such as \textit{Education}, \textit{Healthcare}, and \textit{Art and Media} retain relatively high exposure rates even after normalization, suggesting that LLM-related use cases constitute a larger share of discourse in these fields. Furthermore, the side-by-side comparison in Figure~\ref{fig:exposure_comparison} highlights notable differences between Reddit and News data, where user-driven domains like \textit{Art and Media} and \textit{Education} exhibit higher Reddit exposure, while business-related fields are more prominent in News coverage.

\section{Discussion and Conclusion}

\begingroup
\setlength{\parindent}{0pt}

\subsection{Potential Usage of REALM}  
\vspace{-5pt}

\minisection{Analyzing Occupational Usage Patterns.} \textbf{\textsc{REALM}} facilitates an in-depth analysis of LLM use cases across various occupations, including education, healthcare, and finance. For example, within the education sector, LLMs are frequently discussed for tasks like ``lesson planning, quiz creation, rubric design, and administrative support,''\footnote{\url{https://www.reddit.com/r/edtech/comments/1d676dc/genai_and_education/}} helping teachers reduce preparation time and dedicate more attention to student interactions. Some highlight the need to teach students ``how to ask the important questions,'' as effectively using AI requires clear problem formulation and critical thinking skills that many students still need to develop. Some discussed academic integrity as ``students now rewrite AI outputs with multiple LLMs to make it appear human,''\footnote{\url{https://www.reddit.com/r/Teachers/comments/1d6hnqe/how\_to\_detect\_if\_ai\_wrote\_an\_essay\_when\_the/}} illustrating new challenges. New education methods are developed by redesigning assignments, such as requiring in-class writing or creating AI-focused tasks to foster responsible usage. 
This detailed information complements frameworks like \textit{Particip-AI} \cite{Mun2024ParticipAI}, which evaluates AI risks and benefits through workshops and surveys, by providing large-scale, real-world cases.

\minisection{Studying Bias and Fairness in LLM Applications.} \textbf{\textsc{REALM}} serves as a valuable resource for examining bias and fairness in LLM use cases. Unlike existing datasets that are designed for specific tasks using counterfactual inputs, biased prompts, or masked token evaluation \cite{Gallegos2024Bias}, the biases present in \textbf{\textsc{REALM}} emerge naturally from user-generated content. 
For instance, biases like religious stereotyping and racial profiling are directly reported by users during their engagement with LLM-based systems. 
\minisection{Tools for Developers: Building Robust and User-Aligned LLMs.} 
Our dataset also provides practical value to LLM developers in two key areas: enhancing robustness and improving user alignment. 
\textbf{REALM} offers information about how LLMs are utilized in real-world contexts, thus allowing developers to anticipate trends and proactively design protection mechanisms for evolving vulnerabilities. Additionally, \textbf{REALM} captures firsthand user feedback, providing insights into usability issues and emerging needs. 

\endgroup

\subsection{Conclusion}
This paper introduces \textbf{\textsc{REALM}}, a novel dataset for analyzing real-world LLM use cases across professional domains. It contains data from Reddit and news articles, providing a comprehensive view of LLM usage while acknowledging limitations such as uneven representation. We aim to advance the understanding of LLMs' evolving societal roles by this public dataset.

\section*{Limitations}

While \textbf{\textsc{REALM}} provides valuable insights into the utilization of LLMs across various professions and tasks, it is subject to several limitations. Firstly, the dataset is exclusively derived from Reddit and news articles, excluding other significant platforms such as Twitter, Facebook, and LinkedIn. Secondly, while we attempted to mitigate occupational group imbalance through a debiasing analysis (Section~\ref{sec:debiasing}), our method remains a simplification. Future work could explore more nuanced normalization strategies, such as incorporating engagement signals or semantic relevance.

\section*{Acknowledgments}

This work was supported by funding from Open Philanthropy.


\bibliography{reference}

\appendix

\section{Appendix}
\label{sec:appendix}

\subsection{Taxonomy Selection}
\label{sec:taxonomy selection}

For data labeling in \textbf{REALM}, we select two taxonomies to provide a comprehensive understanding of the use cases. The LLM Use taxonomy categorizes LLM applications based on human goals and outcomes, while the Occupation taxonomy, adapted from the O*NET database, maps these use cases to specific job roles.

The LLM Use taxonomy, as proposed in \citep{Theofanos2024AITaxonomy}, categorizes LLM applications based on human goals and outcomes. This taxonomy is selected due to its strong foundation in real-world AI use cases derived from industry, making it highly relevant for our dataset. It was developed through a rigorous process of refining common AI use activities, balancing abstraction with specificity, and validating its coverage through multiple AI research repositories, including NIST’s AI Community of Interest (COI) Project Catalogue. Although the taxonomy is not specific to technological techniques, its structure is well-suited for categorizing large language model relevant cases. We make minor adjustments to better fit our data labeling task, ensuring it effectively classifies the use cases in our dataset.

For the Occupation taxonomy, we opt for the O*NET database \footnote{\url{https://www.onetonline.org/}}, which is widely recognized for its authority and regularly updated to ensure its timeliness. While we consider alternative resources, such as Indeed, we find that their taxonomy levels do not align well with our data, leading to too much variability and making them unsuitable for the labeling task. O*NET provides additional granularity in occupation levels, however, when we attempt to apply this more detailed classification, we encounter difficulties in extracting sufficient information from reddit posts or news articles to support more specific occupational labels. This resulted in poor labeling consistency, and many finer occupation categories were not adequately covered. A more detailed approach could potentially require case studies, workshops, or additional resources, which we leave for future work.

\subsection{RoBERTa Fine-tuning Details}
\label{appendix:roberta-finetuning}

\paragraph{Model and Hyperparameters.}
We fine-tuned the \texttt{roberta-base} model using the following hyperparameters:

\begin{itemize}[leftmargin=1.5em, itemsep=1pt, topsep=2pt]
  \item Learning Rate: $2 \times 10^{-5}$
  \item Batch Size: 16
  \item Number of Epochs: 10 (applied separately to the two stages of transfer learning)
  \item Optimizer: AdamW
  \item Class Rebalancing: Applied to mitigate class imbalance, with a weight of 3.0 assigned to the minority class (use cases)
  \item Loss Function: CrossEntropyLoss with class weights
  \item Maximum Sequence Length: 512
\end{itemize}

\paragraph{Training Data.}
The model was trained on a total of \textbf{2,000 annotated data points}, comprising 1,000 news articles and 1,000 Reddit posts. Each data point was manually labeled as either a use case or a non-use case. Notably, only approximately 15\% of the samples were labeled as use cases, resulting in a significant class imbalance. To address this, we applied class rebalancing techniques during training to ensure effective learning from the underrepresented class.

The training procedure consisted of two phases: an initial joint fine-tuning stage using the combined dataset, followed by separate fine-tuning on the news and Reddit domains to account for their domain-specific characteristics.

\paragraph{Evaluation.}
The model was evaluated on a held-out test set of \textbf{400 instances}, equally split between 200 news articles and 200 Reddit posts. Despite the low prevalence of use cases in the data, the model achieved a \textbf{high recall score} and a relatively \textbf{high precision score}, demonstrating its effectiveness in identifying use cases under imbalanced conditions.

\begingroup
\setlength{\parindent}{0pt}

\subsection{Details of the Four-Module Annotation Pipeline}
\label{sec:4-modules}
Below are the descriptions of the four modules in the data annotation phase shown in Figure \ref{fig:pipeline}.

\minisection{Summarization Module} generates summaries for each datapoint, which include news articles or Reddit posts along with their top five comments. The summaries focus on describing the LLM use case and the user professions. Additionally, this step filters out irrelevant cases from the high-recall dataset.

\minisection{Classification Module} leverages GPT-4o-mini with few-shot examples to categorize datapoints. Based on task definitions, category explanations, representative examples, and the summaries produced by the Summarization Module. The module also outputs a confidence score indicating the LLM's certainty regarding the generated label, facilitating the identification of classifications that may require further review.

\minisection{Reflection Module} reassesses and refines classifications for datapoints that receive low-confidence scores from the Classification Module. By reflecting on the previously assigned label using the title, summary, and initial classification, the Reflection Module generates a revised label.

\minisection{Multi-Label Linking Module} establishes connections between occupations and LLM usage types through pairwise mappings based on contextual information. This module determines which usage types correspond to specific occupations, enabling a detailed understanding of LLM adoption across various professional domains.

\endgroup

\subsection{Data Annotation Details}
\label{sec:annotation details}
The data annotation process consists of two tasks that serve the data collection and data annotation phase separately. Two expert annotators collaboratively labeled the data, achieving a high inter-annotator agreement (IAA).

The first task involves annotating 2,000 data points selected from the set of Reddit posts and news articles after keyword filtering. We provide clear guideline on what LLM use cases mean to human annotators, details can be found in Appendix~\ref{appendix:annotation-guideline}. These data points were categorized into “Use Case” and “Non-Use Case” labels to train a RoBERTa model, which acted as an automated classifier, thus optimizing resource use and time efficiency. The IAA for this task reached 0.89, and the resulting RoBERTa model achieved high recall scores, as discussed in Section \ref{sec: data_collection}. 

The second task focuses on a subset of 1,000 data points, selected from those identified as ‘Use Case’ after the RoBERTa model’s filtering process, with 500 points from news articles and 500 from Reddit posts. The data points are labeled for “LLM Use Categorization” and “Occupation Categorization” to validate the accuracy of our data annotation pipeline. This task yields IAA scores of 0.87 for LLM Use and 0.85 for Occupations.

Through expert supervision and a context-aware pipeline, the quality of our dataset is ensured.

\subsection{Annotation Guideline for Use Case Classification}
\label{appendix:annotation-guideline}

We provided the following guideline on what LLM use cases mean to human annotators (with extensive examples, though not all are listed here):

\begin{itemize}
  \item \textit{Current Use of LLM: Cases that describe or mention the direct use of large language models (LLMs), including through APIs or products integrated with LLMs that are being used by end-users. These cases always involve clear descriptions of users, or the users themselves are the subjects of the articles, indicating that LLMs or products, software, tools, or systems built on LLMs are being actively used.} \\
  \textit{Example: customer service chatbots used by customers, LLM-based writing tools, or developer-facing APIs. Any direct interaction with an LLM to generate content or provide functionality is also categorized here.}
  
  \item \textit{Potential Use of LLM: Cases that involve the introduction, release, or advertisement of products, tools, or systems built using LLMs, which are mostly speculative or have not yet been adopted by end-users. This excludes ethical discussions or sentiment analysis without actual use.} \\
  \textit{Example: announcements of upcoming LLM-powered features or promotional materials for unreleased tools.}
\end{itemize}

We differentiate between current and potential use cases because the two types of posts or articles may vary significantly. Current use cases usually have clearly defined end-users, while potential use cases often lack clear end-user involvement and may simply describe a newly released product. This distinction helps reduce confusion during the annotation process.

\lstset{
    backgroundcolor=\color{gray!10},   
    basicstyle=\ttfamily\scriptsize,    
    frame=single,                       
    rulecolor=\color{gray},             
    breaklines=true,                    
    captionpos=b,                       
    xleftmargin=0pt,                    
    aboveskip=0pt,                      
    columns=fullflexible,               
    keepspaces=true,                    
}

\begingroup
\setlength{\leftskip}{0pt}

\subsection{Prompts}
%

In the data annotation phase, we use GPT-4o-mini for labeling. Below are the prompts for Reddit posts (the prompts for news articles are similar, with minor adjustments made to accommodate the data characteristics and structure). The content within the curly braces in the prompts represents the input for different data points. We only include the prompts for the Summarization and Single-Stage Classification modules, as they are most critical for understanding and applying our taxonomy.

The Summarization Module not only serves to generate summaries but also improves the precision score of our LLM use case identification (compensating for the sacrifice made by the RoBERTa model to achieve a higher recall score). The definition of use cases in Listing 1 has been independently verified and achieved a precision score of 98\% on the validation set.
\vspace{10pt}

\begin{lstlisting}[caption={Prompts for Summarization Module}]
SYSTEM
You are an expert in analyzing and summarizing use cases of large language models. You will be provided with the title, content and top 5 comments of a reddit post. Based on the human goals and outcomes, summarize in detail how LLM is used and who the end users are, paying attention to their profession or role (including comments). If the human goals or occupation information of end users is not clear, just say you don't know. Your summary should integrate both the usage and occupation information clearly in at least 3 sentences. 

If you think this is not a use case of large language model, just return the phrase 'not a use case' without other words. 
These two senerios should all be considered a use case:
1.Cases that describe or mention large language models being directly used, including through APIs, or products integrated with large language models being used by end-users.
2.Cases that include the introduction, release, or advertisement of products/software/tools/systems developed on top of large language models, mostly speculative or experimental and may not yet have end-user adoption.

USER
Subreddit:  {subreddit}
Title: {title}
Content: {content}
Top_comments: {top_comments}

Please provide a summary explaining both how the LLM is used and detailing the occupation or roles of the end users who benefit from or are expected to use the LLM. Use at least 3 sentences, do not make assumptions. Or if you think this is not a use case of large language model, return the phrase 'not a use case' without other words. 
\end{lstlisting}

In Listing 2, the definition of each LLM use category is modified based on descriptions in the original paper to achieve better classification results.
\vspace{10pt}
\begin{lstlisting}[caption={Prompts for Single-Stage Classification Module (LLM Use)}]
SYSTEM
You are an expert assistant specializing in analyzing and categorizing how LLM is used in a reddit post with its top comments. Here is a large language model use case summary of a reddit post and its comments focusing on end-users and how users use them. Based on the human goals and outcomes of the usage process of LLM stated in the summary, please use the following taxonomy to categorize them. Analyze based solely on the text provided, without making any assumptions, such as infering usage for certain senarios or occpations. 

1. Content Creation: The LLM generates new and original content such as narrative, software code, synthetic data. This involves creativit where the LLM acts as a creator or producer.

2. Content Synthesis: The LLM combines and/or summarizes parts, elements, or concepts into a coherent whole. This involves integrating diverse pieces of information to form a complete and comprehensive output.

3. Decision Making: The LLM assists in selecting a course of action from among possible alternatives to arrive at a solution. This involves evaluating options, predicting outcomes, and recommending the best course of action based on data analysis and algorithms.

4. Detection and Monitoring: The LLM identifies the existence or presence of something through careful analysis and observation, and monitors processes, quality, or states over time. This involves continuous or periodic assessment to gain insights and ensure standards.

5. Digital Assistance: The LLM acts as a personal assistant, understanding and responding to commands and questions, and performing requested tasks in a conversational manner. This involves interaction with users through natural language processing and task execution.

6. Discovery: The LLM aids in finding, recognizing, or uncovering something new. This involves exploratory analysis and innovation to reveal new insights, patterns, or entities.

7. Image Analysis: The LLM processes and interprets textual descriptions related to images to extract meaningful information. This involves recognizing patterns, objects, and relevant details within visual data.

8. Information Retrieval Or Search: The LLM efficiently finds and retrieves relevant information on specific topics of interest. This involves searching large databases, documents, and online resources to provide accurate and relevant data.

9. Personalization: The LLM customizes content, services, or products to match an individual's preferences, behaviors, or characteristics. This involves tailoring user experiences based on interaction history and personal data.

10. Prediction: The LLM forecasts the likelihood of future outcomes based on historical data and trends. This involves using statistical models and machine learning algorithms to predict future events or behaviors.

11. Process Automation: The LLM automates repetitive tasks, removes bottlenecks, reduces errors, and increases efficiency. This involves streamlining processes and saving time by automating routine activities.

12. Recommendation: The LLM suggests or proposes a manageable set of viable options to aid decision-making. This involves analyzing user preferences and context to offer personalized recommendations.
   
13. Robotic Automation: The LLM is used to assist or control physical machines to automate, improve, or optimize tasks. This involves combining LLM capabilities with robotics for complex physical tasks.
   
14. Vehicular Automation: The LLM supports or manages the automation of transportation systems. This involves using LLMs in conjunction with autonomous vehicles for navigation, safety, and decision-making.
    
15. Unknown: This category refers to cases where the LLM's role or specific functionality is ambiguous or not provided. While it's clear that an LLM is used, there is insufficient information to classify its use into any of the known categories.  Notice, This category is mutually exclusive with others.

Here are two examples you can refer to:
Example 1: xxx; Reason: xxx
Example 2: xxx; Reason: xxx

USER
Subreddit: {subreddit}
Article summary: {summary}
The output format should be as follows, do not generate other extra things:
category: list the categories fit the datapoint, separate them with "/", such as : Content Synthesis/Personalization.
explanation: explain the reason for your classification.
confidence: set the confidence to 1 if you are confident in your judgment. If the information is insufficient or you cannot make a reliable judgment, set the confidence to 0.

\end{lstlisting}

In Listing 3, the items listed under "this group includes" are based on the finer-grained occupational divisions from the O*NET database.
\vspace{10pt}
\begin{lstlisting}[caption={Prompts for Single-Stage Classification Module (Occupation)}]
SYSTEM
You are an expert assistant specializing in analyzing and categorizing the current or potential end-user occupation characteristics. Here is a large language model use case summary of a news article focusing on end-users and how users use them. According to the summary and title, please classify the summary based on the definitions and keywords of the following occupational groups. Notice: focus on the end user rather than developers or roles of LLM itself, and do not make assumptions. A summary can include various types of end users. Some have specific occupations, while others are less defined. If a data point indicates a specific occupation, provide that occupation instead of labeling it as "Unknown." Labeling it as "Unknown" implies that the other labels are invalid.
Definitions for each occupation group are as follows:
    
1. Management:
This group includes occupations that involve planning, directing, and coordinating business activities at different levels of management.  
This group include:
	1.	Top Executives
	2.	Advertising, Marketing, Promotions, Public Relations, and Sales Managers
	3.	Operations Specialties Managers
	4.	Other Management Occupations
Keywords: Leadership, Strategic planning, Resource allocation, Policy development

2. Business and Financial Operations:
This group includes occupations that focus on business operations and financial management. People in these roles are involved in analyzing business strategies, managing financial risks, ensuring compliance, and improving organizational efficiency. 
This group include:
	1.	Business Operations Specialists
	2.	Financial Specialists
Keywords: Risk management, Compliance, Financial analysis, Business strategy

3. Computer and Mathematical:  
This group includes occupations related to computer science, IT, and mathematics. People in these fields are engaged in software development, network management, data analysis, cybersecurity, and mathematical research.
This group include:
	1.	Computer Occupations
	2.	Mathematical Science Occupations
Keywords: Software development, Data analysis, Cybersecurity, Mathematical modeling

4. Architecture and Engineering:  
This group includes occupations that focus on the design, construction, and maintenance of buildings, infrastructure, and machinery. Notice: this does not include building ai models, engineers here are experts in real estate.
This group include:
	1.	Architects, Surveyors, and Cartographers
	2.	Engineers
	3.	Drafters, Engineering Technicians, and Mapping Technicians
Keywords: Building design, Infrastructure, Technical plans, Safety standards

5. Life, Physical, and Social Science    
This group includes all scientific researches such as biology, physics, chemistry, and social sciences. Professionals in this category are researchers who contribute to diverse areas, including social science analysis, environmental conservation, behavioral analysis, and public policy development.
This group includes:
	1.	Life Scientists
	2.	Physical Scientists
	3.	Social Scientists and Related Workers
	4.	Life, Physical, and Social Science Technicians
Keywords: Science, Scientific research, Environmental studies, Behavioral science, Policy development, Experimentation

6. Community and Social Service:
This group includes occupations focused on providing social services, counseling, and community support. If cases are just indicating a community-oriented perspective, then it should not fall in this category.
This group include:
	1.	Counselors, Social Workers, and Other Community and Social Service Specialists
	2.	Religious Workers
Keywords: Social work, Counseling, Community support, Social change

7. Legal:
This group includes occupations related to the legal system, where professionals provide legal representation, support judicial processes, and interpret laws. These roles require a strong understanding of legal principles, analytical thinking, and excellent communication skills.
This group include:
	1.	Lawyers, Judges, and Related Workers
    2.  Legal Support Workers
Keywords: Legal representation, Judicial process, Law interpretation, Advocacy

8. Education, Training, and Library:
This group includes occupations that focus on education, training, and providing access to information. These people are responsible for teaching, training, and assisting others in gaining knowledge and skills. Students are also included in this category.
This group include:
	1.	Postsecondary Teachers
	2.	Preschool, Primary, Secondary, and Special Education School Teachers
	3.	Other Teachers and Instructors
	4.	Librarians, Curators, and Archivists
    5.  Other Educational Instruction and Library Occupations
Keywords: Teaching, Curriculum development, Educational resources, Information access

9. Arts, Design, Entertainment, Sports, and Media:
This group includes content creators, writers and editors, and also occupations in creative fields such as design, performing arts, social media, and entertainment.
This group include:
	1.	Art and Design Workers
	2.	Entertainers and Performers, Sports and Related Workers
	3.	Media and Communication Workers
	4.	Media and Communication Equipment Workers
Keywords: Social media, Artistic design, writers, writers, editors, editors

10. Healthcare Practitioners and Support
This group includes occupations involved in diagnosing, treating, and supporting patient care.
This group includes:
	1.	Health Diagnosing and Treating Practitioners
	2.	Health Technologists and Technicians
	3.	Nursing, Psychiatric, and Home Health Aides
	4.	Occupational Therapy and Physical Therapist Assistants and Aides
	5.	Other Healthcare Practitioners and Support Occupations
Keywords: Diagnosis, Treatment, Patient care, Therapy, Caregiving, Medical expertise, Rehabilitation

11. Sales and Related:    
This group includes occupations involved in selling goods and services, maintaining customer relationships, and managing sales transactions. 
This group include:
	1.	Supervisors of Sales Workers
	2.	Retail Sales Workers
	3.	Sales Representatives, Services
	4.	Sales Representatives, Wholesale and Manufacturing
	5.	Other Sales and Related Workers
Keywords: Sales, Advertising, Customer relations, Product promotion, Negotiation

12. Office and Administrative Support:
This group includes occupations that provide clerical and administrative support to businesses and organizations. People in these roles handle tasks such as scheduling, and financial record-keeping.
This group include:
	1.	Supervisors of Office and Administrative Support Workers
	2.	Communications Equipment Operators
	3.	Financial Clerks
	4.	Information and Record Clerks
	5.	Material Recording, Scheduling, Dispatching, and Distributing Workers
	6.	Secretaries and Administrative Assistants
	7.	Other Office and Administrative Support Workers
Keywords: Clerical work, Customer service, Administrative tasks, Record keeping, Secretarial services, Dispatching, Financial clerks
 
13. Production: 
This group includes occupations related to the manufacturing and production of goods. The roles include operate machinery, manage production processes, and ensure quality control in various industries. 
This group include:
	1.	Supervisors of Production Workers
	2.	Assemblers and Fabricators
	3.	Food Processing Workers
	4.	Metal Workers and Plastic Workers
	5.	Printing Workers
	6.	Textile, Apparel, and Furnishings Workers
	7.	Woodworkers
	8.	Plant and System Operators
	9.	Other Production Occupations
Keywords: Manufacturing, Quality control, Machinery operation, Production processes

14. Others:
If categorized in this class, the occupation of the end users is clearly defined and should be one of these categories: Protective Service; Food Preparation and Serving Related; Building and Grounds Cleaning and Maintenance; Personal Care and Service; Farming, Fishing, and Forestry; Construction and Extraction; Installation, Maintenance, and Repair; Transportation and Material Moving; Military Specific.
If it doesn't fit into above occupational categories, you should categorize it into 15. Unknown, not 14. Others

15. Unknown:
This category is used for occupations that cannot be clearly identified within previous groups. It should only be used if there is insufficient information in the article to determine any specific occupation for the end user, or if the LLM's usage is so broad that it does not pertain to a particular occupational context. 

Here are two examples you can refer to:
Example 1: xxx; Reason: xxx
Example 2: xxx; Reason: xxx

USER
Subreddit: {subreddit}
Article summary: {summary}
The output format should be as follows, do not generate other extra things:

category: list the categories fit the datapoint, separate them with "/", such as : Management/Business and Financial Operations.
explanation: explain the reason for your classification.
confidence: Set the confidence to 1 if you are confident in your judgment. If the information is insufficient or you cannot make a reliable judgment, set the confidence to 0.

\end{lstlisting}





\endgroup

\end{document}